\documentclass{article}
\usepackage{smc}
\usepackage{times}
\usepackage{ifpdf}
\usepackage[english]{babel}
\usepackage{cite}
\usepackage{hyphenat}

\def\papertitle{F0 ANALYSIS OF GHANAIAN POP SINGING REVEALS PROGRESSIVE ALIGNMENT WITH EQUAL TEMPERAMENT OVER THE PAST THREE DECADES: A CASE STUDY}
\def\firstauthor{Iran R. Roman}
\def\secondauthor{Daniel Faronbi}
\def\thirdauthor{Isabelle Burger-Weiser}
\def\fourthauthor{Leila Adu-Gilmore}

\newif\ifpdf
\ifx\pdfoutput\relax
\else
   \ifcase\pdfoutput
      \pdffalse
   \else
      \pdftrue
\fi

\ifpdf 
  \usepackage[pdftex,
    pdftitle={\papertitle},
    pdfauthor={\firstauthor, \secondauthor, \thirdauthor \fourthauthor},
    bookmarksnumbered, 
    pdfstartview=XYZ 
   ]{hyperref}

  \usepackage[pdftex]{graphicx}
  \graphicspath{{./figures/}}
  \DeclareGraphicsExtensions{.pdf,.jpeg,.png}

  \usepackage[figure,table]{hypcap}

\else 
  \usepackage[dvips,
    bookmarksnumbered, 
    pdfstartview=XYZ 
  ]{hyperref}  

  \usepackage[dvips]{epsfig,graphicx}
  \graphicspath{{./figures/}}
  \DeclareGraphicsExtensions{.eps}

  \usepackage[figure,table]{hypcap}
\fi

\hypersetup{
    colorlinks,%
    citecolor=black,%
    filecolor=black,%
    linkcolor=black,%
    urlcolor=black
}

\title{\papertitle}

 \fourauthors
   {\firstauthor} {New York University \\ %
     {\tt \href{mailto:roman@nyu.edu}{roman@nyu.edu}}}
   {\secondauthor} {New York University \\ %
     {\tt \href{mailto:df2322@nyu.edu}{df2322@nyu.edu}}}
   {\thirdauthor} {New York University \\ %
     {\tt \href{mailto:ib2225@nyu.edu}{ib2225@nyu.edu}}}
   {\fourthauthor} {New York University\\ %
     {\tt \href{mailto:leila.adu-gilmore@nyu.edu}{leila.adu-gilmore@nyu.edu}}}

\begin{document}
\capstartfalse
\maketitle
\capstarttrue
\begin{abstract}
Contemporary Ghanaian popular singing combines European and traditional Ghanaian influences. We hypothesize that access to technology embedded with equal temperament catalyzed a progressive alignment of Ghanaian singing with equal-tempered scales over time. To test this, we study the Ghanaian singer Daddy Lumba, whose work spans from the earliest Ghanaian electronic style in the late 1980s to the present. Studying a singular musician as a case study allows us to refine our analysis without over-interpreting the findings. We curated a collection of his songs, distributed between 1989 and 2016, to extract F0 values from isolated vocals. We used Gaussian mixture modeling (GMM) to approximate each song’s scale and found that the pitch variance has been decreasing over time. We also determined whether the GMM components follow the arithmetic relationships observed in equal-tem-pered scales, and observed that Daddy Lumba's singing better aligns with equal temperament in recent years. Together, results reveal the impact of exposure to equal-tem-pered scales, resulting in lessened microtonal content in Daddy Lumba's singing. Our study highlights a potential vulnerability of Ghanaian musical scales and implies a need for research that maps and archives singing styles.
\end{abstract}
\section{Introduction}\label{sec:introduction}

Ghanaian popular music today is the result of a long history of European influences interacting with traditional sty-

\noindent les, employing scales with microtonal variance outside e-
\noindent qual temperament. Throughout the 20th century, Ghana featured hybrid music with complex harmony such as Palm-Wine, Highlife and more recently, Hiplife and A-frobeats \cite{collins:01,adu2015studio}. These musics were influenced by traditional Ghanaian music, including European influences such as Christian hymns, sea shanties and brass band music \cite{collins:01}. 

Traditional Ghanaian music is full of features that set it apart from the Western European tradition. Ghanaian harmony is complex and compelling, yet there is a lot of harmonic information to be examined on scales, pitch sets and tuning. Kofi Agawu argues for transcriptions as a valid means of studying African music \cite{agawu2014representing} in keeping with our methods of music theory analysis. A notable in-depth study is Ampene’s research on {\it nwomkoro}, traditional female A- kan singing, with a leader’s call and the chorus’ harmonized response \cite{ampene2005female}. 
Adu-Gilmore (2017) \cite{adu2017music} has researched Ghanaian harmony for over a decade, and although afro- beats has become a best-selling global phenomenon, the culture around African electronic music has been the focus of academic research but the music theory and harmony is still under-researched. Her research has built upon Ampene’s showing parallel harmonies and step-wise chord progressions, creating tonal centers outside of the major and minor modes of the key, implying lesser-used modes in European styles such as Phrygian, Lydian and Mixolydian \cite{adu2017music}. These harmonic tendencies continued to present-day popular music styles such as Ghanaian hip-hop, Highlife and Afrobeats, combining traditional Ghanaian scales with digital instruments and digital audio workstations (DAWs) \cite{adu2015studio}, which are equal-tempered. Moreover, Moelants and colleagues claim that there is a move in African music towards “more elaborate, equally-tempered scales,” even though African music traditionally has differed from the chromatic, equal-tempered twelve-tone scale \cite{moelants2009exploring}.

The international dominance of Western equal-tempered tuning and tonality endangers traditional scales of the world. Internet music dissemination has increased audiences’ music access globally. Ghanaians’ consumption of music from overseas impacts Ghanaian music production and use of effects, including autotune. Although autotune is mainly a means to tune vocals towards equal-temperament, Cher’s 1998 hit “Believe” \cite{higgins1998believe} very audibly used the autotune effect for artistic purposes, producing an unnatural and robotic effect, later popularized by rappers, notably T-Pain, who even created a commercial microphone with an inbuilt autotune \cite{dean2022geophysics}. According to Ghanaian music producer, Kofi “IAmBeatmeance” Boachie-Ansah, Ghanaian tracks such as Sony Achiba’s “Sony Maba” \cite{achiba2001} used autotune early on. Boachie-Ansah states: “This was a few years after Cher’s “Believe,” when that song was dominating the charts. It had a heavy rotation even here, for years […] Accra has always had a very vibrant nightlife culture driven by a lot of dance music so some of the sensibilities that informed those genres of music found a way of seeping into the popular music of the time” \cite{kofi2023}.

We hypothesize that the exposure to music technology and the internet, over the last three decades, has led to musical scales used in popular Ghanaian singing to be more aligned with equal-tempered scales in artists such as Daddy Lumba. This paper presents a case study with quantitative evidence of a move over the years towards pitches that are more in tune with a discrete scale instead of traditional Ghanaian microtonal features, while also gradually trending towards equal temperament. 

To test this hypothesis, we focus on the music of Daddy Lumba, who has been described by Ghanaian popular music scholar John Collins as “one of the most popular Burger Highlife artists” alongside his brothers, Nana Acheampong and Sarkodie \cite{collins:01}. Daddy Lumba's career spans from the earliest Ghanaian electronic style, Burger Highlife in the late 1980s, to similar present-day musics such as Ghanaian Hiplife \cite{charry:01}. Similar to most Ghanaian popular artists, Daddy Lumba is notoriously reclusive and biographical information in general is slim, mostly reliant on music entertainment websites (at times unreferenced, with no author name included) and {\it wikipedia}. Daddy Lumba’s sings music in popular music genres that prominently feature the use of equal-tempered synthesizers and DAWs.

We curated a dataset of Daddy Lumba recordings to ask whether his singing reflects features that progressively abandon traditional Ghanaian scales with microtonal pitch variance in favor of equal temperament. 
We chose the dataset of a singular musician as a case study to refine our analysis method and to avoid over-interpreting the findings. 
Limiting the scope in this way allows us to ensure specificity in the results.  

By presenting empirical evidence, we argue that Ghanaians’ increased access to equal-tempered music through internet dissemination coupled with greater access to music technology is exponentially speeding up the threat to traditional Ghanaian music intonation and scale variance. That is, equal-tempered synthesizers and effects such as autotune enable a more pronounced tendency toward equal-temperament. Far from combating these phenomena, this case study aims to map this change and call for archiving and preserving traditional global musics and their practice.

\section{Related work}\label{sec:relatedwork}

Although music scales have been examined over centuries, their study using large datasets is made possible by two major advances in music information retrieval (MIR). The first consists of audio source separation algorithms, such as Wave-U-Net \cite{stoller2018wave}, recurrent neural networks \cite{huang2014singing}, and spectrogram U-Nets \cite{jansson2017singing},  which can be used to automatically isolate vocal or single instrument tracks\footnote{these type of algorithms have recently been made available in dedicated software toolboxes like Spleeter \cite{hennequin2020spleeter}}. The second advancement consists of F0 estimation algorithms that operate directly on waveform signals, such as YIN \cite{de2002yin, mauch2014pyin} and CREPE \cite{kim2018crepe}. Together, these advancements allow for the extraction of singing F0 values, and the subsequent approximation of musical scales via statistical modeling. 

Previous research has used F0 estimation methods in non-Western music recordings. A recent study combined source-separation and F0 estimation methods to quantitatively analyze melodic mode switching in raga performances \cite{shikarpurcomputational}. Additionally, the YIN algorithm has been used, on its own, to extract the scales present in African and Turkish instrumental melodies \cite{six2011tarsos,bozkurt2014computational}, as well as Indian singing \cite{serra2011assessing,koduri2012raga}. These studies sought to assess whether the scales found in digital musical recordings with real musical practice are consistent with music theories for each of these music traditions \cite{tzanetakis2014computational}. More recently, researchers used CREPE to compare the pitch content among human signing, speech, and birdsongs, finding that human singing, across cultures and countries, emphasizes integer ratio relationships between pitch content (i.e. perfect fifths), compared to speech or birdsongs \cite{kuroyanagi2019automatic}. Similar analysis (but using YIN instead of CREPE) on child and adult singing was used to study the possible sensorimotor origins of inter-ratio relationships commonly seen in sung pitch content \cite{sato2019automatic}.

These studies have been important in showing how modern MIR tools can be applied to analyze melodic pitch content in digital recordings of music. We have yet to find a study, however, that  uses a data-driven methodology to analyze the impact of equal temperament on Ghanaian scales. We therefore believe that our study will be the first in the field to show such an effect. 

\begin{figure*}[h]
  \includegraphics[width=\textwidth]{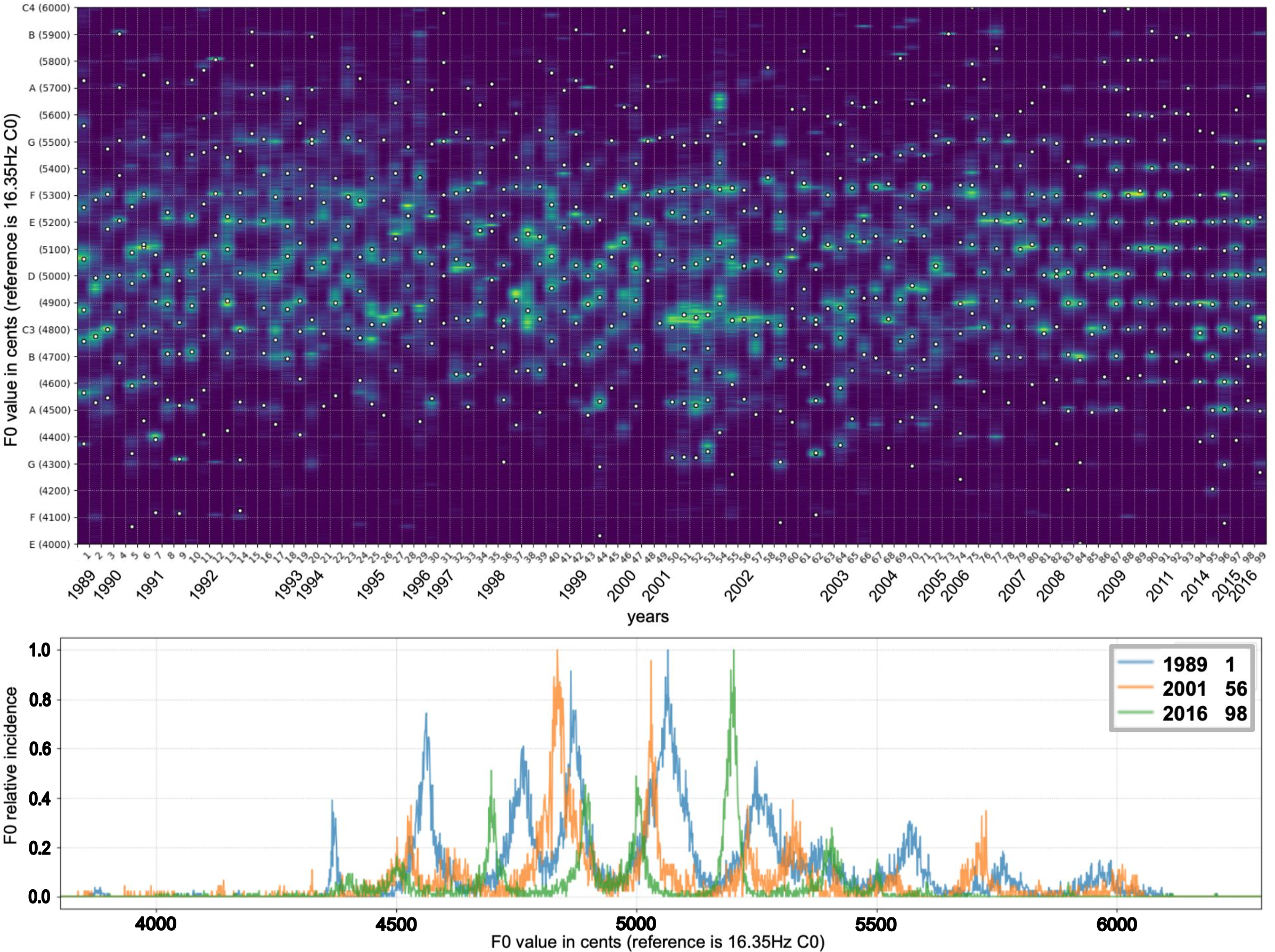}
  \caption{Top panel: each song's F0 histogram, column-wise. Lighter colors indicate higher density. White dots represent the optimal GMM components found. Bottom panel: three sample F0 histograms (normalized so that the F0 with the highest incidence is equal to 1, for visualization purposes).}\label{fig:figure1}
\end{figure*}

\section{Method to estimate F0 values in Daddy Lumba's singing}\label{sec:methods}

Our goal is to study the changes in the scales and pitch content sung by Daddy Lumba over the course of twenty-seven years, as captured in his discography. To do this, we compiled a corpus of 99 songs released between 1989 and 2016 (stereo, mp3, 44.1 kHz sampling rate). Table \ref{table:lumbacorpus1} shows a summary of the most relevant metadata in our corpus. 

\begin{table}
\begin{tabular}{||c c c c||} 
 \hline
 songs per year & song dur & song dur & total  \\ 
 (1989-2016) & (avg) & (std dev) & duration \\ [0.5ex] 
 \hline\hline
 3.54 & 5m42s & 1m8s & 9h23m23s \\ 
 \hline
\end{tabular}
\caption{Column 1: the average number of songs included in our dataset per year. Columns 2 and 3: the average song duration and standard deviation, respectively. Column 4: the total duration of our dataset.}
\label{table:lumbacorpus1}
\end{table}
\begin{table}
\begin{tabular}{||c c c||} 
 \hline
 F0 dur. : song dur. & F0 dur. : song  dur. & total F0\\ 
 (avg.) & (std. dev.) & duration \\ [0.5ex] 
 \hline\hline
 0.22 & 0.06 & 2h5m37s \\ 
 \hline
\end{tabular}
\caption{Columns 1 and 2: the average and standard deviation of the ratio between duration of F0 values extracted and total song duration. Column 3: the total duration of F0 values extracted from our dataset.}
\label{table:lumbacorpus}
\end{table}

Our corpus contains mastered songs, which means that vocals, instruments, and other effects are mixed in the tracks. We use CREPE \cite{kim2018crepe} to do F0 estimation of the vocal tracks. However, CREPE was designed and tested to work on monophonic signals (i.e. tracks with only one instrument) \cite{kim2018crepe}. To isolate the vocals from all other instruments and effects in a song, we used the pre-trained Wave-U-Net made available by its original authors\footnote{https://github.com/f90/Wave-U-Net-Pytorch}. Using the Wave-U-Net on each song, we obtained tracks containing only the vocals sung by Daddy Lumba and vocal accompaniment.

Having the isolated vocal track, we can use CREPE to estimate the F0s in Daddy Lumba’s songs. A pre-trained version of CREPE is made available by its original authors\footnote{https://github.com/marl/crepe}. Given an audio signal, this version of CREPE estimates an F0 value, in Hz, every 10ms, and each value is accompanied by a confidence score between 0 and 1. We used CREPE version v0.0.7 to extract F0 values from each song’s isolated vocal track. To ensure the quality of the F0 values we applied the following heuristic filtering method: 1) we discarded F0 values outside the range between 80Hz and 600Hz (the vocal range of male singing) 2) we also discarded F0 values that were given a confidence score below 0.8 by CREPE. Finally, we converted the F0 values from Hz to cents using \eqnref{cents}  
\begin{equation}\label{cents}
F0_{cents}= 1200 \log_2\bigg(\frac{F0_{Hz}}{16.352_{Hz}}\bigg),
\end{equation}
where 16.352 Hz corresponds to the C0 tone that we used as reference. Table 2 gives a summary of the amount of F0 values that we were able to obtain from the dataset using our extraction and filtering method. In total, we were able to obtain over 2 hours of F0 values. On average, from each song we obtained 22\% of its duration as F0 values.

We used these F0 values to carry out two analyses that quantify changes over Daddy Lumba’s singing career. First, we used a Gaussian mixture model (GMM) to approximate the scale of each song and study the variance around pitches in the sung scale. Second, we used the GMM results to determine whether Daddy Lumba's singing scales reflect an equal-tempered structure.

\section{Analysis 1: estimating song scales and variance around them over years}\label{sec:analysis1}

In this analysis we approximate each song's scale to look at the variations in pitch content. We observe whether this variation has changed over Daddy Lumba's career. Our specific research question is: has Daddy Lumba’s singing reduced its pitch variations over the years? 

\subsection{Methodology}\label{subsec:analysis1method}

\begin{figure*}[h]
  \includegraphics[width=\textwidth]{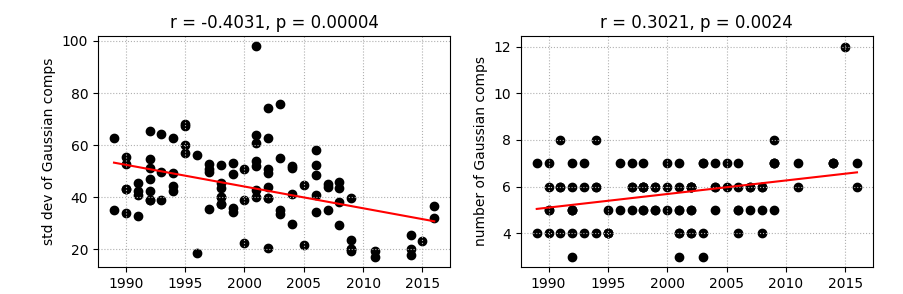}
  \caption{Left panel: a linear regression reveals a significant negative relationship for the F0 GMM variance over the years. Right panel: a linear regression reveals a significant positive relationship for the number of F0 GMM components over the years.}\label{fig:figure2}
\end{figure*}

In general, a Gaussian mixture is a probabilistic unsupervised method that models data as a probability distribution 
that is the sum of different weighted Gaussians. \eqnref{gmm} describes the general equation
\begin{equation}\label{gmm}
p(\theta)= \sum_{i=0}^{C-1} \phi_i\mathcal{N}(\mu_i,\sigma_i),
\end{equation}
where $\mu_i$ and $\sigma_i$ are the $i$'th component's mean and variance, and each component is weighted by a scalar $\phi_i$.

For each song’s F0 values, we estimate the parameters of a Gaussian mixture with “tied” covariance \cite{pedregosa2011scikit}. For the specific case of “tied” covariance, the coefficient $\phi_i=1$ is used and a constant $\sigma$ is found for all components in the GMM \cite{reynolds2009gaussian}. In other words, the data is clustered by using the same distance criterion across the range of F0 values. We use a tied covariance because we are assuming that the distribution of F0 values sung around each pitch in a scale is the same (i.e. Daddy Lumba's singing has a characteristic variability around scale tones, independent of how high- or low-pitched a tone is). We limited the  F0 values considered for the GMM to be an an octave around the most commonly-sung F0 value in each song. 

We validate number of components $C$ in a GMM via the highest euclidean silhouette score around each Gaussian component mean value. The silhouette score is a heuristics based method that yields a coefficient by
\begin{equation}\label{silhouette}
s = \frac{1}{N}\sum_{i=0}^{N-1}\frac{b_i-a_i}{\max(a_i,b_i)},
\end{equation}
where $a_i$ is the average distance between a point $i$ and the other points in the same Gaussian component (i.e. point-wise inter-cluster distance), and $b_i$ is the average distance between a point $i$ and the points in the closest Gaussian component that $i$ is not part of. The silhouette score ranges from -1 to 1, with values close to 1 indicating that datapoints, on average, belong to their assigned Gaussian component, while -1 indicates that datapoints, on average, do not belong to their assigned Gaussian component. Since our GMM components have a ``tied'' covariance, the silhouette score is a fair metric to validate components that should have similar density properties.

We tested integer values between 4 and 15 for the possible number of components, with 4 being the lowest number of pitch classes that we considered to be reasonable for a singing scale, and 15 the maximum within an octave. The GMM with the number of components that maximized the silhouette score was selected to model a song’s scale. 

\subsection{Results: singing scales approximated via GMM}\label{subsec:analysis1result}

The top panel of figure~\ref{fig:figure1} shows a matrix where columns are song histograms of F0 values, chronologically ordered from left to right. White dots are the mean of each component in the song’s F0 GMM. On average, the silhouette score associated with each song’s F0 GMM was 0.72 (lowest was 0.62 and highest was 0.87). The GMM components (i.e. the white dots) are an approximation of each song's scale based on our GMM approach. The bottom panel of Figure~\ref{fig:figure1} shows three example F0 histograms, normalized so that the highest peak of each is equal to 1 (for visualization purposes). Note how the spread around peaks is visually smaller for the years 2016 and 2001 compared to 1989.
With that said, we should be mindful that our dataset has a limited number of data points from 2015 onwards. 
We believe that less data starting at that point is not a big problem since, as shown in the top panel of Figure \ref{fig:figure1}, Daddy Lumba's singing reflects the use of autotune. 

\subsection{Results: F0 GMM components over the years}

The left panel of figure~\ref{fig:figure2} shows the standard deviation associated with each song's F0 GMM components. A linear regression revealed a significant negative correlation (r = -0.3839; p $<$ 0.05), indicating that, as years have progressed, the vocals have reduced their microtonal pitch diversity, suggesting a tendency to sing more “in tune” with the song’s scale. The right panel of figure~\ref{fig:figure2} shows the number of GMM components in each song's F0 GMM. A linear regression revealed a significant positive correlation (r = 0.2834; p $<$ 0.05), indicating that, over the years, Daddy Lumba's sings scales with more tones per octave.  

\section{Analysis 2: Quantifying alignment with equal temperament over years}\label{sec:analysis2}

Analysis 1 resulted in a GMM for each song, with a unique number of components and standard deviation around them, as well as a characteristic distance between components. 

\subsection{Methodology}

In a song that perfectly follows the equal-tempered scale, the distance between any two notes $n_i$ and $n_j$ follows this rule (in units of cents): 
\begin{equation}\label{equal-temp}
|n_i - n_j| = N\times100,
\end{equation}
where 100 is the shortest possible distance between two consecutive notes in a scale, and $N \in \mathbf{Z}$ is the number of times such distance separates the notes $n_i$ and $n_j$. 100 cents is the smallest possible distance between equal-tem- pered notes. However, no singer is perfect and musical performances have an error term associated, making \eqnref{equal-temp}: 
\begin{equation}\label{equal-temp-error}
|n_i - n_j| = N\times100 + \epsilon,
\end{equation}
where $\epsilon$ is a value between 0 and 50, reflecting an error from the perfect minimal distance of 100 cents. An $\epsilon$ closer to zero indicates better alignment with equal-temperament.  

We want to quantify to what extent Daddy Lumba's singing reflects a scale that is consistent with equal temperament, and whether his singing has progressively become more equal-tempered over the years. Hence, we calculated \eqnref{equal-temp-error} between the F0 GMM components found in each song to get the song's average $\epsilon_s$. 

\subsection{Results: equal temperament alignment over years}

\begin{figure}  \includegraphics[width=\linewidth]{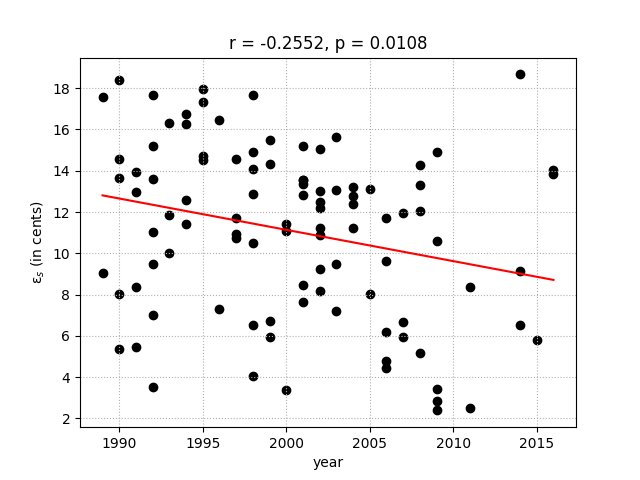}
  \caption{A linear regression of how error $\epsilon_s$ associated with deviations from equal temperament has been reducing over the years, indicating alignment with an equal-tempered scale}\label{fig:figure3}
\end{figure}

A linear regression over the songs' $\epsilon_s$ values revealed a significant negative correlation (r = -0.2552; p $<$ 0.05), indicating that, over the years, Daddy Lumba's singing has progressively reduced the $\epsilon$ error term in \eqnref{equal-temp-error}. Figure~\ref{fig:figure3} shows this regression. 

The code and F0 data needed to reproduce our results and generate the figures in this paper is openly available\footnote{https://github.com/iranroman/AFRINUM}.

\section{Discussion}\label{sec:discussion}

We are interested in examining the microtonal nuances in Ghanaian singing and its associated musical scales. This initial study analyzes Daddy Lumba’s work over time because he has an influential catalog of work spanning from early Ghanaian electronic music called Burger Highlife in the 90s, until the present moment. 

The first analysis used GMM to approximate the singing scales in Daddy Lumba’s songs (independent of whether they are equal-tempered or not), and looked closely at the pitch variations. The results in the left panel of figure~\ref{fig:figure2} showed that, over time, he has tended to sing closer to the discrete tones in his pitch sets (or modes), with less microtonal variance (i.e. he sings more “in tune” with the scale he uses in a song). The potential reasons for this could be manifold. For instance, his personal experience and developing artistic identity could be the driving force, as Daddy Lumba's career has spanned many years, and this experience could mean that, over time, the precision of his singing has been refined to target his chosen scale. However, over the same time-span, the pop music industry has seen an increased access to high-quality, budget alternatives to recording studios, such as DAWs that allow singers to iterate over potential recordings and select tracks where their voice reflects their preferred tuning precision and scale. Editing in DAWs is typically a lot faster, and different takes may be easier to record, and edit \cite{bell2018dawn}. These changes in technology have implications everywhere.

We also analyzed how Daddy Lumba sings in relation to equal temperament, and exposed a gradual shift towards equal temperament over the years (see Figure~\ref{fig:figure3}). The reasons for this may also be linked to access to technology. First, traditional scales may be affected by fixed pitch equal-tem- pered instruments, such as piano, sound synthesizers, and wind instruments, as the string instruments and voices are tuned to them. Second, electronic instruments and software using “auto-tuned” MIDI are tuned to the equal-tempered system. Third, music made using electronic instruments and computers is pervasive. Fourth, the mass music distribution through the internet has increased the speed at which equal-tempered popular music is disseminated, consumed and absorbed. Therefore, digital instruments and effects embedded with tonal harmony of Wes- tern-European Christian scales and equal temperament may impact Ghanaian singing, as well as other music across the globe with traditional scales, tuning conventions and vocal inflections. 

Why do we compare Daddy Lumba's singing against equal temperament? Because of its history as a British colony, to study Ghanaian popular music today, Daddy Lumba included, it is necessary to analyze it through the lens of both Western European tonal harmony and traditional Ghanaian harmony. Critical Sonic Practice (CSP) investigates global music by dissolving borders between music technology, composition and theory, while including indigenous epistemologies. Western-European Christian tonal harmony is necessary in this discussion, not because this scale is “in tune,” but because these scale systems are a cultural choice at the heart of the European colonizing force of music \cite{adu2021embodied}.   

\section{Conclusion}\label{sec:conclusion}

This study aims to catalyze a scientific discourse about how global access to music technology has influenced Gha- naian music, popular and traditional, as well as other global genres, including the global north, where traditional styles, like flamenco in Spain or R\'{i}mur in Iceland, feature microtonal inflections that cannot be defined by the equal-tempered system \cite{gomez2008automatic, ingolfsson2019jon}. Evolution in music formats, such as vinyl to tape dubbing, already increased access to broader genres; whereas internet dissemination increased access to equal-tempered music from abroad, and gives Ghanaians more access to global markets that picked equal temperament as an unofficial standard. Furthermore, recent advances in technology, such as cheaper (and free) DAWs, virtual studio technology (aka VST), and the prevalence of autotune increase the ubiquity of equal temperament. Changes in music scales and temperament are inevitable and MIR is an essential tool in mapping this change. 

The findings in this study align with the mission of CSP, which includes decolonizing datasets, and departing from norms of tuning that are often overlooked, taken for granted, or even unknown. In doing so, we recognize that current algorithms are often trained with datasets that reflect majoritian musics. The specificity of this paper as a case study, which closely examines one artist, signifies that we can more easily understand the culturally-relevant nuances of a given artist’s corpus. This allows us to foresee potential errors in cultural bias or expectation. Therefore, we are fine-tuning necessary steps in research design that are culturally sensitive to the dataset, before continuing to move forward toward our future work with larger datasets encompassing entire genres, music traditions, or geographical regions. 

To reiterate, this paper demonstrates a research design for detecting microtonal variance over time in the music of one artist, rather than a comprehensive study of Ghanaian harmony. That said, the findings point towards more research into Ghanaian modes and tonalities, as well as the changes affected by technologies.

\section{Future Research}\label{sec:future}

We will expand our work by analyzing other artists in the same genre. We also aim to research and expand the timeline to encompass releases before 1989, hoping to better map and describe traditional scales and influences. Future works using MIR could involve an in-depth study of Ghanaian scales and modes in the same dataset or others to expand the archive of Ghanaian popular and traditional musics. These could be cross-referenced with expert listening panels to verify the empirical data. This research will be augmented by interviews with artists, ensuring culturally specific context for the broader study in order to investigate the reason behind this phenomenon. The work presented here lays the groundwork for the empirical, data-driven archiving of traditional and local modes and tonalities. With these goals in mind, this study has broader implications in the decolonization of music research requiring archiving of tonal systems from all over the world for cultural preservation. 


\bibliography{smc2023bib}

%
%
%
%
%

\end{document}